\documentclass[aps,prd,twocolumn,10pt,groupedaddress]{revtex4-1}
\usepackage{amssymb}
\usepackage{graphicx}
\usepackage{amsmath}
\usepackage{hyperref}
\usepackage{subfigure}
\usepackage{float}
\usepackage[utf8]{inputenc}

\begin{document}

\title{Paper-boat relaxion}
\author{Shao-Jiang Wang}
\email{schwang@cosmos.phy.tufts.edu}
\affiliation{Tufts Institute of Cosmology, Department of Physics and Astronomy, Tufts University, 574 Boston Avenue, Medford, Massachusetts 02155, USA}

\begin{abstract}
A new relaxion mechanism is proposed where a small electroweak scale is preferably selected earlier than the larger one due to a potential instability, which is different from previously proposed stopping mechanisms by either Hubble friction from an increasing periodic barrier or thermal friction from gauge boson production. The sub-Planckian field excursion of an axion can be achieved without violating the bound on the $e$-folding number from a quantum gravity perspective; and our relaxion can be identified as a QCD axion, preserving the Peccei-Quinn solution to the strong $CP$ problem as well as making up all the cold dark matter in our current Universe.
\end{abstract}
\maketitle

\section{Introduction}

The electroweak (EW) hierarchy problem of the standard model (SM) remains as a major challenge after the null search at the Large Hadron Collider (LHC), inspiring many new ideas in addition to the traditional approaches from supersymmetry \cite{Dimopoulos:1981zb}, extra dimensions \cite{Randall:1999ee,ArkaniHamed:1998rs}, and strong dynamics \cite{Weinberg:1975gm,Susskind:1978ms}. Among these new ideas, the so-called relaxion mechanism \cite{Graham:2015cka} has received much attention recently. The essential idea behind the relaxion mechanism is to dynamically relax the electroweak scale via cosmological evolutions. Similar to the cosmological relaxation of the EW scale during inflation in the relaxion mechanism, the recent idea of $N$naturalness \cite{Arkani-Hamed:2016rle} also solves the EW hierarchy by the cosmological selection of the EW scale during reheating. (See also \cite{Geller:2018xvz,Cheung:2018xnu} for the cosmological selection of the EW scale during inflation.)

The potential in the relaxion mechanism for the Higgs $h$ and axion $a$ contains three parts, $V(a,h)=V_\mathrm{Higgs}+V_\mathrm{scan}+V_\mathrm{stop}$, of the form
\begin{align}
V_\mathrm{Higgs}&=\frac{\lambda}{4}h^4-\frac12(\Lambda_h^2-\Lambda_h^2\frac{a}{f})h^2;\\
V_\mathrm{scan}&=c_1\Lambda_h^4\frac{a}{f}+c_2\Lambda_h^4\left(\frac{a}{f}\right)^2+\cdots;\\
V_\mathrm{stop}&=\Lambda_a(\langle h\rangle)^4\left(1-\cos\frac{a}{f_a}\right).
\end{align}
The Higgs part $V_\mathrm{Higgs}$ contains an axion-dependent mass-square, triggering EW symmetry breaking (EWSB) when it scans from a large cutoff scale $\Lambda_h$ to a critical point at $a=f$ where it becomes negative. The scanning part $V_\mathrm{scan}$ could generally include the higher-order terms in $a/f$ that gradually become unimportant once the axion rolls down to $a<f$ \cite{Patil:2015oxa}. The dimensionless positive parameter $c_n$ is of order unity from the naturalness argument \cite{Choi:2016luu}. Note that $V_\mathrm{stop}$ quickly stops the axion from further rolling due to Higgs backreaction soon after it passes through $a=f$, where the growing barrier amplitude $\Lambda_a^4=\Lambda_c^{4-n}\langle h\rangle^n$ can  either be generated from QCD dynamics $(a/32\pi^2f)G\widetilde{G}$ with $n=1$ or other hidden strong dynamics with $n=2$ \cite{Espinosa:2015eda}. Here, $\Lambda_c$ is the strong coupling condensation scale of some gauge group with gauge field strength $G$. The current vacuum expectation value (VEV) $v$ of the Higgs is selected from the stopping condition in which the slope from the linear scanning term balances the slope from the periodic potential at the current axion value $a_0$, namely,
\begin{align}
\left(\frac{v}{\Lambda_h}\right)^n\sim\frac{f_a}{f}\left(\frac{\Lambda_h}{\Lambda_c}\right)^{4-n}\frac{1}{\sin(a_0/f_a)}.
\end{align}
Therefore, the relaxion mechanism solves the EW hierarchy problem by trading the EW hierarchy $v\ll\Lambda_h$ with the axion hierarchy $f_a\ll f$ \cite{Choi:2016luu}, which is technically natural since those terms involving $f$ in the potential that explicitly break shift symmetry of the axion should be highly suppressed by sufficiently large enough $f$.

Despite the elegance in the original relaxion mechanism \cite{Graham:2015cka}, there are some unusual features for model building. For the scanning part, one generally expects a super-Planckian field excursion of the axion with a gigantic $e$-folding number during ultra low-scale inflation. These requirements come from slowly scanning enough Higgs mass range so that it is independent from initial conditions during the slow-roll era dominated by classical evolutions other than quantum fluctuations. For the stopping part, if the relaxion is a QCD axion, one obtains an EW cutoff scale as well as an inflation scale below QCD confinement, and the final misalignment angle of the QCD vacuum is of order unity, which destroys the Peccei-Quinn (PQ) solution to the strong $CP$ problem. See \cite{Nelson:2017cfv,Jeong:2017gdy,Davidi:2017gir}, however, for alternative realizations of the QCD relaxion.

Besides the dynamical stopping mechanism by Hubble friction from an increasing periodic barrier as a result of backreaction of the Higgs after EWSB, there is an alternative stopping mechanism by thermal friction from gauge particle production as a result of an axion coupling to weak gauge bosons \cite{Hook:2016mqo,Fonseca:2018xzp} (see also \cite{Choi:2016kke,Choi:2016kke,Matsedonskyi:2017rkq,Tangarife:2017vnd,Tangarife:2017rgl}). The energy dissipation during gauge particle production is most efficient when the Higgs VEV is small; hence, a small EW scale is thus selected. The primary motivation for this thermal stopping mechanism is to eliminate the need for an ultra-low-scale inflation with a gigantic $e$-folding number over a super-Planckian field excursion. However, the idea of the axion coupling only to weak gauge bosons but not to photons might be lost at higher-order corrections where dissipation to photons could stop the axion from approaching the small EW scale. See \cite{You:2017kah,Son:2018avk} for alternative realizations of the relaxion with particle production.

In this paper, we propose a new relaxion mechanism with a stochastic stopping mechanism that is different from the dynamical or thermal stopping mechanisms we mention above. The current EW scale is naturally selected due to the potential instability. Furthermore, a consistent choice of parameters can be identified to render a QCD relaxion as all of the cold dark matter (CDM).

\section{General picture}

\begin{figure}
\centering
\includegraphics[width=0.5\textwidth]{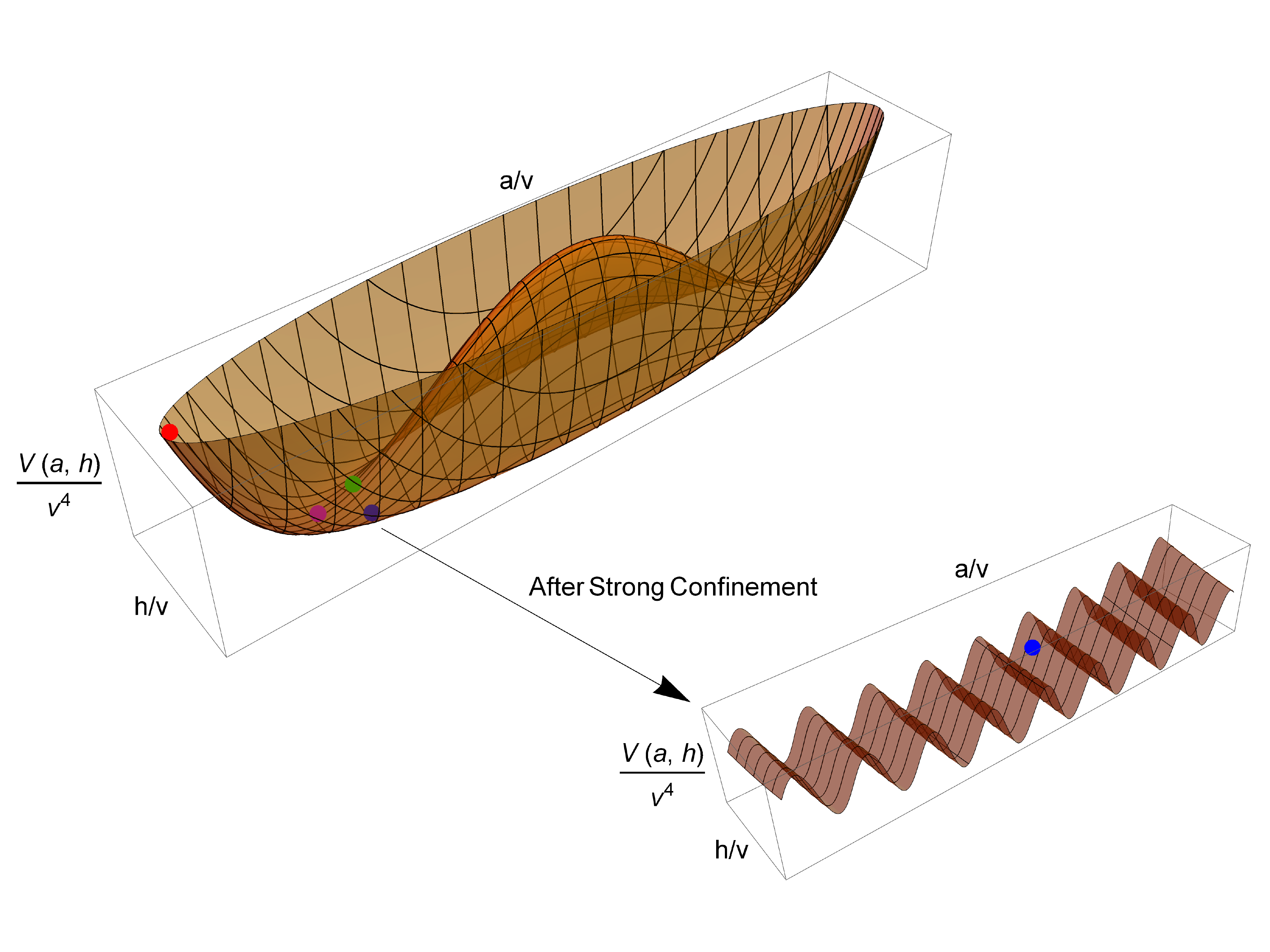}\\
\caption{Sketch of the paper-boat relaxion mechanism. The red point is the release point where the relaxion starts rolling. The magenta point is the critical point where EW symmetry starts break. The green point is the deviation point where Higgs fluctuations kick in. The blue point is the relaxation point where the Higgs is eventually relaxed to its current VEV, which is securely locked by the periodic potential of the axion soon after strong confinement. For clarity, the potential is not to scale, and it should be highly squeezed along the Higgs axis direction and extremely stretched along the axion axis direction.}\label{fig:MexicanRelaxion}
\end{figure}

Our new relaxion mechanism admits a Mexican-hat-like potential
\begin{align}\label{eq:MexicanRelaxion}
V(a,h)&=\frac{\lambda}{4}\left(h^2-\frac{\Lambda_h^2-ga^2}{\lambda}\right)^2=\frac{1}{4\lambda}(ga^2+\lambda h^2-\Lambda_h^2)^2\nonumber\\
&=\frac{\lambda}{4}h^4+\frac12(ga^2-\Lambda_h^2)h^2+\frac{\lambda}{4}\left(\frac{\Lambda_h^2-ga^2}{\lambda}\right)^2
\end{align}
where $\Lambda_h$ is the cutoff scale, and $g$ is a technically natural small coupling that explicitly breaks the shift symmetry of the axion $a$. We include the appearance of a periodic potential $\Lambda_a^4(1-\cos(a/f_a))$ below the confinement scale $\Lambda_c$ with axion decay constant $f_a$. The quadratic coupling in the scanning term of the Higgs mass was adopted before in \cite{Huang:2016dhp}. The sketch of this Mexican-hat-like potential is presented in Fig.\ref{fig:MexicanRelaxion}, where all the local minimums are degenerated along an ellipse,
\begin{align}\label{eq:ellipse}
\frac{a^2}{\Lambda_h^2/g}+\frac{h^2}{\Lambda_h^2/\lambda}=1.
\end{align}
Since $g$ is generally much smaller than $\lambda$, this Mexican-hat-like potential is therefore highly squeezed along the Higgs axis direction and extremely stretched along the axion axis direction, hence the name of paper-boat relaxion mechanism as shown in Fig.\ref{fig:MexicanRelaxion}.

The general picture of our paper-boat relaxion mechanism is very simple. Since the Higgs VEV is zero before EWSB, the axion can be released along the axion axis direction from any point $a\gtrsim\mathcal{O}(1)\Lambda_h/\sqrt{g}$ (called the release point, shown as a red point in Fig.\ref{fig:MexicanRelaxion}). The rolling axion will scan the mass-square of the Higgs from $\mathcal{O}(1)\Lambda_h^2$ down to zero until passing through $a=\Lambda_h/\sqrt{g}$ (called the critical point and specified by a magenta point in Fig.\ref{fig:MexicanRelaxion}), where EW symmetry starts break. It is obvious that the axion axis direction is stable against Higgs fluctuations in the range $a>\Lambda_h/\sqrt{g}$ before EWSB, while it is unstable against Higgs fluctuations in the range $a<\Lambda_h/\sqrt{g}$ after EWSB. Therefore, the relaxion certainly cannot climb up the ridge further away from the critical point until Higgs fluctuations kick in at some point (called the deviation point and specified by a green point in Fig.\ref{fig:MexicanRelaxion}) where relaxion falls off the hillside to a point (called the relaxation point and specified by a blue point in Fig.\ref{fig:MexicanRelaxion}) in the valley of the ellipse; thus, the current VEV of the Higgs is accidentally selected due to the potential instability, and it is further securely locked by the periodic potential of the axion below the confinement scale. 

The general picture above is lacking specific background cosmological evolution, for which each stage will be constrained below.

\section{Scanning stage}
 
Suppose that the critical point where EW symmetry starts break is around $T_\mathrm{EW}\sim10^2\,\mathrm{GeV}$ within the radiation dominated era; thus, the leading-order thermal corrections should not be important below the critical point to destroy the valley of the ellipse, and the scanning stage from the release point to the critical point covers all the background inflationary era up to the current observable scale, as well as the reheating era and part of the radiation dominated era. To be independent of the initial condition, the part of the scanning stage during inflation with $a\sim\mathcal{O}(1)\Lambda_h/\sqrt{g}$ should be slow rolling,
\begin{align}\label{eq:slowroll}
m_a^2(h=0)\simeq\mathcal{O}(1)\frac{g}{\lambda}\Lambda_h^2\ll H_\mathrm{inf}^2\Rightarrow g\Lambda_h^2\ll\lambda H_\mathrm{inf}^2.
\end{align}
Meanwhile, the classical evolution should dominate the quantum fluctuations in the first place,
\begin{align}\label{eq:classical}
\left.\frac{V'_a}{3H_\mathrm{inf}^2}\right|_{h=0}>\frac{H_\mathrm{inf}}{2\pi}\Rightarrow\sqrt{g}\Lambda_h^3\gtrsim\lambda H_\mathrm{inf}^3,
\end{align}
and the energy density of the relaxion must not disturb the background inflation,
\begin{align}\label{eq:inflation}
V(a,h=0)\ll M_\mathrm{Pl}^2H_\mathrm{inf}^2\Rightarrow\Lambda_h^4\ll\lambda M_\mathrm{Pl}^2H_\mathrm{inf}^2.
\end{align}
In addition, the field excursion of the axion should also be large enough for sufficient scanning,
\begin{align}\label{eq:scanning}
\Delta a\simeq\left.\frac{V'_a}{3H_\mathrm{inf}^2}\right|_{h=0}\Delta N\simeq\mathcal{O}(1)\frac{\Lambda_h}{\sqrt{g}}\Rightarrow\Delta N\simeq\frac{\lambda H_\mathrm{inf}^2}{g\Lambda_h^2}.
\end{align} 
Note that two extra constraints from quantum gravity can be imposed, namely, a sub-Planckian field excursion \cite{Giddings:1987cg} and a not-too-large $e$-folding number \cite{ArkaniHamed:2007ky},
\begin{align}\label{eq:QGbound}
\Delta a\lesssim M_\mathrm{Pl},\quad \Delta N\lesssim M_\mathrm{Pl}^2/H_\mathrm{inf}^2.
\end{align}
With the help of \eqref{eq:classical} for a sub-Planckian inflationary scale, 
\begin{align*}
\Delta a\simeq\frac{\sqrt{g}\Lambda_h^3}{\lambda H_\mathrm{inf}^2}\Delta N\lesssim M_\mathrm{Pl}
\Rightarrow\Delta N\lesssim\frac{\lambda H_\mathrm{inf}^2M_\mathrm{Pl}}{\sqrt{g}\Lambda_h^3}\lesssim\frac{M_\mathrm{Pl}^2}{H_\mathrm{inf}^2},
\end{align*}
a sub-Planckian axion excursion \eqref{eq:QGbound} automatically preserves the $e$-folding bound \eqref{eq:QGbound}.
Combining \eqref{eq:slowroll} with \eqref{eq:classical} leads to $g^2\ll\lambda$. Combining \eqref{eq:classical} with \eqref{eq:inflation} leads to $\lambda H_\mathrm{inf}^6\ll g^2M_\mathrm{Pl}^6$. Therefore, $g$ is constrained by
\begin{align}\label{eq:gbound}
\sqrt{\lambda}(H_\mathrm{inf}/M_\mathrm{Pl})^3\ll g\ll\sqrt{\lambda}.
\end{align}
Combining \eqref{eq:classical} with $g^2\ll\lambda$ leads to $\Lambda_h^3\gg g^{3/2}H_\mathrm{inf}^3$. Therefore, $\Lambda_h$ is constrained by
\begin{align}\label{eq:cutbound}
g^\frac12H_\mathrm{inf}\ll\Lambda_h\ll\lambda^{1/4}M_\mathrm{Pl}^{1/2}H_\mathrm{inf}^{1/2}.
\end{align}
It turns out that, we could meet all the constraints above by taking
\begin{align}\label{eq:choice1}
\begin{split}
g\simeq10^{-4}, \Lambda_h\simeq10^{15}\,&\mathrm{GeV}, H_\mathrm{inf}\simeq10^{14}\,\mathrm{GeV},\\
\Delta N\simeq\mathcal{O}(10)&, \Delta a\simeq10^{17}\,\mathrm{GeV}.
\end{split}
\end{align}
Therefore, our paper-boat relaxion mechanism could naturally avoid the super-Planckian field excursion during ultra-low-scale inflation with a gigantic $e$-folding number that usually plagues the original relaxion mechanism \cite{Graham:2015cka}, and our cutoff scale could be as large as the grand unification theory (GUT) scale. However, as we will see later, the naive choice \eqref{eq:choice1} cannot survive the selecting and securing stages, which impose further constraints in order to eliminate any fine-tuning.

\section{Selecting stage}

When the axion passes through the critical point, the scanning mass term of the Higgs in \eqref{eq:MexicanRelaxion} becomes negative; namely, the EW symmetry starts break. After that, the relaxion continues to climb up the ridge, which is unstable against either quantum or thermal fluctuations along the Higgs direction. After a small field excursion $\delta a$ below the critical point $a=\Lambda_h/\sqrt{g}$, the relaxion is accidentally kicked off the ridge by fluctuations along the Higgs direction, and it eventually rolls down the hillside to a local minimum in the ellipse $g(\Lambda_h/\sqrt{g}-\delta a)^2+\lambda\langle h\rangle^2=\Lambda_h^2$. The obtained VEV of the Higgs is therefore technically small,
\begin{align}\label{eq:vev}
\frac{\langle h\rangle^2}{\Lambda_h^2}\sim\frac{\sqrt{g}}{\lambda}\frac{\delta a}{\Lambda_h}\ll1,
\end{align}
due to a technically small $g$ or a naturally small hierarchy $\delta a\ll\Lambda_h/\sqrt{g}$, which is another way of saying that the deviation point $\langle a\rangle=\Lambda_h/\sqrt{g}-\delta a$ where the Higgs fluctuations kick in should be very close to the critical point. However, the relaxed Higgs VEV $\langle h\rangle$, or equivalently $\delta a$, cannot be predicted as an established value because one simply cannot predict exactly when Higgs fluctuations kick in and the relaxion rolls down the hillside, whereas it most likely falls off the ridge soon after it passes through the critical point. In fact, the probability of the relaxion climbing up the ridge without being kicked down the hillside is exponentially suppressed from a crude estimation below. If the probability for the Higgs to stay on the ridge at some time ($e$-folding number) $N$ is $P(N,h=0)$, then the probability to find the Higgs still on the ridge after some time ($e$-folding number) $\delta N$ can be solved by the Fokker-Planck equation
\begin{align}\label{eq:FPeq}
\frac{\partial P}{\partial N}=\frac{\partial}{\partial h}\left(\frac{V'_h}{3H^2}P+D^2\frac{\partial P}{\partial h}\right).
\end{align}
If the Higgs classically slow rolls down the hillside over the fluctuation $D$ term after an accidental kick (and eventually obtain a VEV $\langle h\rangle$), then a crude estimation reads
\begin{align}\label{eq:prob}
\frac{P(N+\delta N,h=0)}{P(N,h=0)}\simeq1+\frac{m_h^2}{3H^2}\approx\mathrm{e}^{-\frac{\lambda\langle h\rangle^2}{3H^2}\delta N}.
\end{align}
Therefore, the probability to stay on the ridge is exponentially suppressed because the would-be Higgs VEV becomes larger and larger when climbing up the ridge; hence, a technically and accidentally small Higgs VEV is expected. However, it actually becomes more and more difficult to slow roll down the hillside; thus, the slow-roll approximation made in \eqref{eq:FPeq} only serves as a conservative estimation, and we expect the qualitative conclusion to remain unchanged for a more rigorous analysis \cite{Bartrum:2014fla,Cheung:2015iqa}. 

For our naive choice \eqref{eq:choice1}, one expects $\delta a\sim10^{-10}\,\mathrm{GeV}$ for an EW-scale Higgs VEV $\langle h\rangle\sim10^2\,\mathrm{GeV}$. Such a small $\delta a$ is actually fine-tuned because, if the relaxion climbs up the ridge further by even a small amount of $\delta a\sim10^{-9}\,\mathrm{GeV}$, then the corresponding deviation in the Higgs VEV would be as large as $\delta h=(g\langle a\rangle/\lambda\langle h\rangle)\delta a\simeq10^3\,\mathrm{GeV}$, resulting in an even smaller probability of staying there. In this view, it is reasonable to expect such seemingly fine-tuned $\delta a$. However, thermal fluctuations of order $T_\mathrm{EW}\sim10^2\,\mathrm{GeV}$ at the onset of EWSB could easily bump the relaxion up along the ridge so as to shift the relaxed Higgs VEV by an amount of $\delta h=(g\langle a\rangle/\lambda\langle h\rangle)\delta a\simeq10^8\,\mathrm{GeV}$. To avoid this problem, one necessarily encounters a suspicious coincidence problem of why the relaxion rolls down the hillside even before thermal fluctuations are ever developed along the ridge. Fortunately, one can easily acquire a larger $\delta a$ at the price of allowing for a larger $e$-folding number of inflation at a lower scale, which can be seen from combining \eqref{eq:scanning} with \eqref{eq:vev}, namely,
\begin{align*}
\frac{\lambda H_\mathrm{inf}^2}{\Delta N\Lambda_h^2}\lesssim g\lesssim\frac{\lambda^2v^4}{\delta a^2\Lambda_h^2}\Rightarrow\delta a^2\lesssim\frac{\lambda v^4}{H_\mathrm{inf}^2}\Delta N\ll\frac{\lambda v^4M_\mathrm{Pl}^2}{H_\mathrm{inf}^4}.
\end{align*}
Hence, an EW scale $\delta a\sim10^2\,\mathrm{GeV}$ requires an inflationary scale below $10^{10}\,\mathrm{GeV}$. With $\langle h\rangle\sim10^2\,\mathrm{GeV}$, one has
\begin{align}\label{eq:choice2}
\begin{split}
g\simeq10^{-16},\,&\Lambda_h\simeq10^9\,\mathrm{GeV},\,H_\mathrm{inf}\simeq10^6\,\mathrm{GeV},\\
\Delta N\simeq10^{9},\,&\Delta a\simeq10^{17}\,\mathrm{GeV},\,\delta a\sim10^2\,\mathrm{GeV},
\end{split}
\end{align} 
which meets all the constraints we have discussed so far, as well as evading the fine-tuning and coincidence problems of $\delta a$, because thermal fluctuations of order $T_\mathrm{EW}\sim10^2\,\mathrm{GeV}$ now lead to a shift for the Higgs VEV of order $\delta h=(g\langle a\rangle/\lambda\langle h\rangle)\delta a\simeq10^2\,\mathrm{GeV}$, which is sufficient to solve the EW hierarchy problem with the most conservative precision.

\section{Securing stage}

After the relaxion climbs up the ridge until the deviation point and then rolls down the hillside, the Higgs is found in a local minimum $(\langle a\rangle, \langle h\rangle)$ along both directions from the ellipse equation \eqref{eq:ellipse},
\begin{align}
\frac{\partial V}{\partial a}&=g\langle h\rangle^2\langle a\rangle+\frac{g}{\lambda}(g\langle a\rangle^2-\Lambda_h^2)\langle a\rangle=0,\\
\frac{\partial V}{\partial h}&=\lambda\langle h\rangle^3+(g\langle a\rangle^2-\Lambda_h^2)\langle h\rangle=0,
\end{align}
where the evaluation of the partial derivative at $(\langle a\rangle, \langle h\rangle)$ is understood. Since different regions of our Universe could roll down the hillside at different deviation points with the width of thermal fluctuations $T_\mathrm{EW}\sim10^2\,\mathrm{GeV}$, the achieved Higgs VEV also admits some distribution with width $\delta h\sim10^2\,\mathrm{GeV}$ under \eqref{eq:choice2}, leading to an inhomogeneous distribution of Higgs VEVs in our Universe. Furthermore, before the nonperturbative effect at the strong confinement scale $\Lambda_c$ takes place (if it is lower than the EW scale), the drift of VEV due to fluctuations along the flat direction in the valley of the ellipse is of order $\sqrt{\delta a^2+\delta h^2}\simeq\delta a\sim10^2\,\mathrm{GeV}$, which, under \eqref{eq:choice2}, results in a shift of the Higgs VEV of order $\delta h\simeq10^2\,\mathrm{GeV}$, which also leads to an inhomogeneous distribution of Higgs VEVs in our Universe. This can be solved after strong confinement with the presence of the periodic potential, $V_\mathrm{tot}=V+\Lambda_a^4(1-\cos(a/f_a))$, where the inhomogeneous distribution of Higgs VEVs rolls down to the new minimum set by
\begin{align}
\frac{\partial V_\mathrm{tot}}{\partial a}&=\frac{\partial V}{\partial a}+\frac{\Lambda_a^4(\langle h\rangle)}{f_a}\sin\frac{\langle a\rangle}{f_a}=0,\\
\frac{\partial V_\mathrm{tot}}{\partial h}&=\frac{\partial V}{\partial h}+\frac{\partial\Lambda_a^4}{\partial h}\left(1-\cos\frac{\langle a\rangle}{f_a}\right)=0,
\end{align}
namely, a set of discrete points in the original ellipse with interval $\delta a=2n\pi f_a$. The final misalignment angle would be exactly zero at these discrete vacuums, which preserves the PQ solution \cite{Peccei:1977hh,Peccei:1977ur} to the strong $CP$ problem if our relaxion is chosen as the QCD axion. It is worth noting that the exactly vanishing energy density of the axion potential at these discrete vacuums also evades the argument from the de Sitter quantum breaking bound \cite{Dvali:2013eja,Dvali:2014gua,Dvali:2017eba} (see also \cite{Dvali:2018txx,Dvali:2018dce,Ibe:2018ffn} for recent arguments on the existence of the axion). Nevertheless, the way out of the swampland \cite{Brennan:2017rbf,Obied:2018sgi,Ooguri:2018wrx}\cite{Garg:2018reu} at these discrete vacuums should be discussed separately.

However, for the QCD axion with decay constant $f_a\gtrsim10^9\,\mathrm{GeV}$ bounded from below by supernova cooling observations \cite{Brockway:1996yr,Grifols:1996id}, the improved choice \eqref{eq:choice2} would cause another fine-tuning problem that, unless a minimum of the periodic potential is coincided with the relaxed minimum during the selecting stage--namely, a fine-tuning relation $\Lambda_h^2-g(2n\pi f_a)^2\simeq\lambda v^2$ is conspired--the relaxion obtained during the selecting stage would generally roll down the periodic potential to the new minimum with an axionic shift of order $\delta a\sim f_a$.  Since the Higgs VEVs of two adjacent minimums would differ by an amount of $\delta h=(g\langle a\rangle/\lambda\langle h\rangle)f_a\simeq10^9\,\mathrm{GeV}$, this would necessarily destroy the desired solution obtained by the selecting stage. Fortunately, there exists a parameter space for achieving $\delta a$ as large as the decay constant $f_a$ of the QCD axion, which also meets all the constraints we have discussed so far, as well as a desirable Higgs VEV $\langle h\rangle\sim10^2\,\mathrm{GeV}$ with the most conservative precision $\delta h\sim10^2\,\mathrm{GeV}$,
\begin{align}\label{eq:choice3}
&\Lambda_h\simeq\frac{\lambda\langle h\rangle\delta h}{\sqrt{g}\delta a}\simeq10^6\,\mathrm{GeV}\left(\frac{g}{10^{-24}}\right)^{-\frac12}\left(\frac{f_a}{10^9\,\mathrm{GeV}}\right)^{-1},\nonumber\\
&H_\mathrm{inf}\simeq\frac{g^\frac16}{\lambda^\frac13}\Lambda_h\simeq10^2\,\mathrm{GeV}\left(\frac{g}{10^{-24}}\right)^{-\frac13}\left(\frac{f_a}{10^9\,\mathrm{GeV}}\right)^{-1},\nonumber\\
&\Delta a\simeq\frac{\Lambda_h}{\sqrt{g}}\simeq10^{18}\,\mathrm{GeV}\left(\frac{g}{10^{-24}}\right)^{-1}\left(\frac{f_a}{10^9\,\mathrm{GeV}}\right)^{-1},\nonumber\\
&\Delta N\simeq\frac{\lambda H_\mathrm{inf}^2}{g\Lambda_h^2}\simeq10^{15}\left(\frac{g}{10^{-24}}\right)^{-\frac23}.
\end{align}
Requiring a sub-Planckian field excursion $\Delta a\lesssim M_\mathrm{Pl}$ and a cutoff scale $\Lambda_h$ above the TeV scale, one immediately obtains the allowed range of $g$ as $10^{-15}/f_a\lesssim g\lesssim1/f_a^2$, which is small if the decay constant of the axion is large.

Now that the width of the deviation point is much smaller than the decay constant, all the relaxed Higgs VEVs will be distributed within the same period of the periodic potential closest to the critical point, and will eventually settle down at the same minimum below the confinement scale; thus, there is no inhomogeneous problem for the Higgs VEVs. The current Higgs VEV can also be achieved with a precision better than $\delta h\simeq10^2\,\mathrm{GeV}$ for a decreasing product $\sqrt{g}\Lambda_h$. Our solution to the EW hierarchy problem is thus accomplished here.
 
\section{QCD relaxion DM}

It turns out that the parameter space of \eqref{eq:choice3} also embraces an appealing choice of $f_a\simeq10^{11}\,\mathrm{GeV}$ with
\begin{align}\label{eq:choice4}
\begin{split}
\Lambda_h&\simeq10^4\,\mathrm{GeV}\left(\frac{g}{10^{-24}}\right)^{-1/2},\\
H_\mathrm{inf}&\simeq1\,\mathrm{GeV}\left(\frac{g}{10^{-24}}\right)^{-1/3},\\
\Delta a&\simeq10^{16}\,\mathrm{GeV}\left(\frac{g}{10^{-24}}\right)^{-1},\\
\Delta N&\simeq10^{15}\left(\frac{g}{10^{-24}}\right)^{-2/3}.
\end{split}
\end{align}
We find that our QCD relaxion could also make up all CDM \cite{Bae:2008ue} without fine-tuning the initial misalignment angle \cite{Preskill:1982cy,Abbott:1982af,Dine:1982ah} and without violating the current upper bound \cite{Akrami:2018odb} on the isocurvature perturbation. In this case, our QCD relaxion mass is $2g^2a^2/\lambda\simeq2g\Lambda_h^2/\lambda\sim10^{-15}\,\mathrm{GeV}$ before confinement and $10^{-14}\,\mathrm{GeV}$ after confinement \cite{Weinberg:1977ma,Wilczek:1977pj}. One can also  constrain $10^{-26}\lesssim g\lesssim10^{-22}$ and the cutoff scale $1\,\mathrm{TeV}\lesssim\Lambda_h\lesssim10^2\,\mathrm{TeV}$. 

Note that both $e$-folding numbers in \eqref{eq:choice3} and \eqref{eq:choice4} satisfy \eqref{eq:QGbound}, which can be further reduced by considering a realistic inflationary background \cite{Patil:2015oxa} with Hubble flow parameter $\epsilon_H=-\dot{H}/H^2\simeq\mathcal{O}(10^{-2})$, namely,
\begin{align}
\Delta N\simeq\log\left(1+3\epsilon_H\frac{\lambda H_\mathrm{inf}^2}{g\Lambda_h^2}\right)^{1/\epsilon_H}\simeq\mathcal{O}(10^3).
\end{align}
Also note that we do not expect any gravitational waves from inflation at the observable level. However, the required low-scale inflation is unusual for realistic model-building. Nevertheless, the current bound on the inflationary Hubble scale is quite loose. The upper bound $H_\mathrm{inf}\lesssim10^{13}\,\mathrm{GeV}$ comes from the Lyth bound \cite{Lyth:1998xn} $H_\mathrm{inf}=1.06\times10^{-4}r^{1/2}M_\mathrm{Pl}$ with the tensor-to-scalar ratio $r<0.064\,(95\%\,\mathrm{CL})$ at the pivot scale $k_*=0.002\,\mathrm{Mpc}^{-1}$ from the combination of the Planck 2018 and the BICEP2/Keck Array BK14 data \cite{Akrami:2018odb}, or $r<0.06\,( 95\%\,\mathrm{CL} )$ at $k_*=0.05\,\mathrm{Mpc}^{-1}$ from BICEP2/Keck Array BK15 data \cite{Ade:2018gkx}. The lower bound on the inflationary Hubble scale is only theoretically required to be at least $H_\mathrm{inf}\gtrsim T_\mathrm{reh}^2/M_\mathrm{Pl}\sim10^{-14}\,\mathrm{GeV}$ for a successful reheating temperature $T_\mathrm{reh}\gtrsim10^2\,\mathrm{GeV}$. Lower reheating temperatures down to the nucleosynthesis scale are also allowed by observation. Please refer to, e.g., Ref. \cite{German:2001tz,Postma:2004an,Dimopoulos:2004yb,Rodriguez:2004yc,BuenoSanchez:2007jxm,Allahverdi:2009rm,Ross:2010fg,Bramante:2016yju,Guth:2018hsa} for specific model-building of low-scale inflation also in the context of the relaxion mechanism \cite{Evans:2017bjs}.

It is also interesting to explore the possibility of generating the baryon asymmetry of our Universe (BAU) by baryogenesis with a QCD axion \cite{Kuzmin:1992up,Servant:2014bla,DeSimone:2016bok,Jeong:2018ucz,Jeong:2018jqe} or by leptogenesis via Higgs relaxation \cite{Kusenko:2014lra,Pearce:2015nga,Yang:2015ida} during the selecting stage.

\section{Conclusions and discussions}

A new relaxion mechanism is proposed in the most economic manner that selects our current EW scale with a stochastic stopping mechanism, which is different from previous proposed mechanisms, either dynamical or thermal stopping. Not only is a small EW scale naturally obtained, but a comparable precision is also achieved; therefore, no fine-tuning is needed for our new relaxion mechanism. Without violating the bounds from quantum gravity on the field excursion and $e$-folding number, a consistent choice of parameters can be identified for our QCD relaxion that not only preserves the PQ solution to the strong $CP$ problem but also makes up all the CDM in our current Universe.

However, we point out two crucial issues of the form of our effective potential \eqref{eq:MexicanRelaxion}. First, such an effective potential should be derived from a more fundamental UV theory. It is worth noting that a very similar picture, sharing the same local potential shape around the critical point, was adopted in a recent paper \cite{Amin:2019qrx}, where the Higgs is coupled to a scalar modulus with a specific SUSY realization. This might point to a possible direction for our UV completion. Second, the paper-boat structure of our effective potential could be radiatively unstable against quantum corrections unless it is protected by some symmetries at low energies. One such symmetry admitted by our effective potential is a rotationlike symmetry,
\begin{align}
\left(\begin{matrix}a'\\h'\end{matrix}\right)=
\left(\begin{matrix}
\cos\theta & \sqrt{\lambda/g}\sin\theta\\
-\sqrt{g/\lambda}\sin\theta & \cos\theta
\end{matrix}\right)
\left(\begin{matrix}a\\h\end{matrix}\right),
\end{align}
which, unfortunately, is not shared by the kinetic term unless the rotation angle $\theta$ is a Grassmann number after taking the real part of the kinetic term. Furthermore, this rotationlike symmetry is not sufficient to uniquely fix the form of our effective potential. Additional terms involving $g a^2+\lambda h^2$ as a whole can be added into our effective potential. Both issues of UV completion and radiative stability go beyond the scope of the current paper, and they merit further study in the future.

\begin{acknowledgments}
We are grateful to Sebastian Bramberger, L.H. Ford, Mark Hertzberg, Mudit Jain, Gustavo Marques-Tavares, Yi Wang, Yi-Peng Wu, and Masaki Yamada for their useful and stimulating discussions and correspondences. This work is supported by funds from Tufts university.
\end{acknowledgments}

\bibliographystyle{utphys}
\bibliography{ref}

\end{document}